\documentclass[final,5p,times,twocolumn]{elsarticle}
\usepackage{amssymb}
\usepackage{lineno}
\usepackage[utf8]{inputenc}
\usepackage[T1]{fontenc}
\usepackage{epsfig}
\usepackage{amsmath}
\usepackage{amsfonts}
\usepackage{amssymb}
\usepackage{color}
\usepackage{graphicx}
\usepackage{url,hyperref}
\usepackage{latexsym}
\usepackage{longtable}
\usepackage{float}
\usepackage{epstopdf}
\usepackage{url}
\usepackage{caption}

\begin{document}
\begin{frontmatter}
\title{\rightline{\mbox{\small
{08-2018}}}\textbf{\ Magnetic Monopoles from a Hidden Magnetic
Symmetry}}
\author{Adil Belhaj $^{a}$}
\author{Salah Eddine Ennadifi $^{b}$}
\author {  \corref {mycorrespondingauthor}}
\cortext[mycorrespondingauthor] {Corresponding author}
\ead{ennadifis@gmail.com}
\address{
$^{a}$ {ERPTM,  Physics department, Polydisciplinary Faculty, Sultan
Moulay Slimane
University,  Beni Mellal, Morocco.}\\
$^{b}$ {Independent Researcher, Rabat, Morocco.} }

\begin{abstract}
It is proposed that Magnetic Monopoles (MMs) could originate from a new $%
\mbox{U(1)}_{M}$  symmetry. Such an abelian symmetry is then assumed
to be related to the conservation of a magnetic number $M$. This
magnetic number is associated with massive MMs from an expected high
scale breaking of  this magnetic symmetry. The involved scales are
approached and the properties of such MMs are investigated along
with the prospects for their detection.

\end{abstract}

\end{frontmatter}
%


\section{Introduction}

Recently, with the significant improvement of cosmological,
astrophysical and neutrino observation quality \cite{1,2,3,4}, the
search for Magnetic Monopoles (MMs) becomes increasingly important
to probe a modern physics going beyond de the Standard Model of
particle physics (SM) \cite{4,5,6,60}. The existence of MMs is
theoretically motivated from the fact that they are predicted by
most non trivial high energy theories including Grand Unified
Theories (GUT) and String Theory\cite{7,8,9,10}. It has been
suggested that they could be exploited to discuss many long-standing
problems in relations  with the Dark Matter (DM) origin remaining a
mystery in both particle physics and cosmology \cite{7,10,11}.
Although the origin  of MMs in the Universe is theoretically
supported, their experimental evidence is a great challenge in the
particle physics community \cite{3,4,5,6}. Concretely, the study of
the properties of MMs is then among the most thrilling problem in
modern physics since they are considered as a fascinating
possibility opening a new road towards a new physics above the SM
\cite{9,10,12}. In the SM, gauge singlets could, in principle, play
a primordial role in the elaboration of the physics dealing with
MMs. In particular, this is practically possible because they do not
involve physical properties associated with the color and the
electroweak gauge anomalies of fundamental interactions. Assuming
that they are neutral under some new gauge forces, their magnetic
number is not constrained. In fact, the most suggested patterns of
MMs turn out to be difficult to prove or refute by using direct
observations. The simplest method, but well incited pattern, could
be based on a possible extension of the SM by introducing additional
magnetic charged particles. They have specific couplings to the
physics of the SM visible sector. In this way, definite and proper
predictions of the MMs sector should be provided in terms of certain
scalars, the involved symmetry, and the consequences of its
spontaneous breaking.

In this work,  we build a simple physical model in which MMs are
considered as particles whose existence belongs to a high scale
magnetic symmetry. Using a simple abelian symmetry $\mbox{U(1)}_{X}$
extension of the SM, we show how MMs $m$ can easily be accommodated.
Precisely, we examine in some details the breaking of such a
magnetic symmetry along with the associated magnetic number $X=M$
predicting a natural origin of a potential MM candidate. Then, we
deal with massive MMs $m$ and we reveal how this can emerge from the
breaking scale of the proposed magnetic symmetry by supporting the
case of heavy MMs. We also speculate on the searches for MMs.

This paper is organized as follows. In section 2, we list some
theoretical arguments in favor to  MMs. In section 3, we provide a
simple model in which a new magnetic symmetry $\mbox{U(1)}_{M}$
extending the SM and its breaking generate MMs. The searches for MMs
are discussed in section 4. The last section is devoted to a summary
and outlook.

\section{Theoretical arguments for MMs}

We start by recalling that the large part of the observable Universe is made
of electrically charged elementary particles being  quarks $%
q_{f=u,d,c,s,t,b}$, some leptons $\ell _{f=e,\mu ,\tau }$, and
certain gauge bosons $W^{\pm }$. Up to now, though no magnetic
charged elementary particle has been observed, there are various
theoretical reasons for believing that the MMs must exist in nature.
Let us, therefore, cute some arguments being emerged in the
discussion on the existence of MMs supported by many
pioneering investigation works \footnote{%
Throughout this work natural units $c=\hbar =1$ are used.}:

\begin{enumerate}
\item The first argument is based, essentially, on the asymmetry between the
magnetic and the electric fields in Maxwell equations. This problem
remains an open issue in the related  physics \cite{7}.

\item The observed quantization of the electric charge can be seen as a
possible evidence for the existence of MMs \cite{9,13,14,10,15}.

\item It has been suggested that the magnetic charge of a monopole could be
considered as a topological charge associated with various topologies of
physical field configurations \cite{150}. In this way, the correspondence
magnetic charge/topology can be investigated from gauge group theory.

\item MMs can be arisen at spontaneous symmetry breaking of certain non
abelian symmetries associated with particular  topological
properties of some manifolds \cite{151,152}.

\item It has been proposed  that most high-scale theories including GUT of
extended SM  models involving large gauge symmetries should contain MMs \cite%
{7,8,9,10}.

\item String theory also predicts that MMs must exist as solitons which
could be discussed in terms of D-brane physics \cite{153}.

\item The problem of MMs is one of the three cosmological motivations,
flatness and horizon, in favor to cosmic inflation \cite{11}.

\item A confirmed fact is that SM is just a low-energy limit of a more
extended theory \cite{16,17,18,19}. There must be an  undiscovered
physics surpassing the SM scale which can be explored to unveil
certain properties of MMs.
\end{enumerate}

It is important to recall that the magnetic charge $M\ $of MMs relys on the
elementary electric charge $e$ according to the Dirac basic relation \cite{9}
as

\begin{equation}
M=\frac{n}{2e}=nM_{D},  \label{1}
\end{equation}%
where $n$ is an integer and $M_{D}$ is the Dirac unit of the magnetic charge
being
\begin{equation}
M_{D}=\frac{1}{2e}=\frac{e}{2\alpha _{e}},  \label{2}
\end{equation}%
where $\alpha _{e}=e^{2}\simeq 1/137$ is the fine structure constant. In an
analogous manner, a dimensionless magnetic coupling constant is defined as
\begin{equation}
\alpha _{M}=M_{D}^{2}\gg 1.  \label{3}
\end{equation}%
Assuming that it is larger than $1$, this magnetic coupling constant
prevents reliable perturbative calculations of MM production processes.
Thus, according to Dirac theory in which the fundamental charge is $e$, the
minimum value of the magnetic charge is $M_{D}$. However, it would become
multiples of $M_{D}$, in particular $2M_{D}$ or $3M_{D}$, as shown in
various works \cite{9,13,14,10,15}.

Based on these theoretical arguments, one should support the idea that there
is no particular reason to evince the investigation of the existence and the
origin of MMs. From a particle physics view, one should encourage the search
for such particles and shed lights on their hidenness in nature. In what
follows, we discuss this physical problem in the context of a SM extension
involving an abelian symmetry associated with MMs relying on the $\mbox{U(1)}%
_{M}$ group and extra physical fields.

\section{Magnetic extension model}

In this section, we build the model from a simple extension of the SM by
introducing an extra abelian continuous symmetry $\mbox{U(1)}_{X}$ and a
single complex field $S$. It has been shown that there are many roads to
handle such physical pieces. Properly, it is noted that models beyond the SM
can be engineered using theories involving extra dimensions and
supersymmetry including superstring and M-theories. In this regards, the
geometry and the topology of extra dimensions engender abelian gauge
symmetries $\mbox{U(1)}^{\prime }$s and scalar fields $S_{i}$ using
different approaches either in the context of M-theory on $G2$ manifolds or
in intersecting type II D-branes wrapping non trivial cycles embedded in
Calabi-Yau manifolds \cite{154,155,156,157}. It has been revealed that the
scalar fields can play an important role in the elaboration of new physical
models extending the SM. Assuming that the present model could be obtained
from such stringy inspired theories, we consider this extended gauge
symmetry of the SM
\begin{equation}
G_{SM+MMs}=G_{SM}\otimes \mbox{U(1)}_{X}.  \label{4}
\end{equation}%
The first group refers to  the SM symmetry $G_{SM}=\mbox{SU(3)}_{C}\otimes %
\mbox{SU(2)}_{L}\otimes \mbox{U(1)}_{Y}$ describing three generations of
quarks and leptons governed by the strong, weak and electromagnetic
interactions \cite{16,17,18,19}. It has been confirmed that the
corresponding theory is an interesting quantum field theory model of
elementary particles and fundamental interactions in nature. The additional
group $\mbox{U(1)}_{X}$ is associated with the conservation of a quantum
number $X$ to be specified later on. It is noted in passing that this
symmetry could find a place in string theory, M-theory, and related models
based on D-brane objects.

In the present model (\ref{4}), all the SM  fields are assumed to be
neutral under this new symmetry. This can be ensured by the
constraint
\begin{equation}
X\left( q_{i}\right) =X\left( \ell _{i}\right) =X\left( H\right) =0,
\label{5}
\end{equation}%
where   the fields $q_{i}$ and $\ell _{i}$ are the SM quarks and
leptons, respectively and  $H$ represents the SM Higgs field. The
subscript ($i=1,2,3$) refers to the three generations of SM
fermions.  The simplest way allowing for the breaking of this
symmetry is to impose the new symmetry $\mbox{U(1)}_{X}$ on the
complex scalar field $S$. This means that this scalar becomes
charged under such a symmetry as required by
\begin{equation}
X\left( S\right) \neq 0.  \label{6}
\end{equation}%
However, this scalar field $S$ is neutral under the SM symmetry
$G_{SM}$ such that
\begin{equation}
C\left( S\right) =I\left( S\right) =Y\left( S\right) =0\text{.}  \label{7}
\end{equation}%
A careful examination shows that, for this field-extended SM, the most
general renormalizable scalar potential takes the following form
\begin{equation}
V(H,S)_{SM+S}=V(H)_{SM}+\mu _{S}^{2}|S|^{2}+\lambda _{S}|S|^{4}+\lambda
_{SH}|S|^{2}|H|^{2}.  \label{8}
\end{equation}%
Here, the first term $V(H)_{SM}$ $=\mu _{H}^{2}|H|^{2}+\lambda _{H}|H|^{4}$ denotes the usual Higgs potential of the SM where the field $%
H^{T}=(h^{0},h^{-})$ is the standard Higgs doublet of the $\mbox{SU(2)}_{L}$
symmetry. The parameters $\mu _{H}^{2}$, $\lambda _{H}$, $\mu _{S}^{2}$, $%
\lambda _{S}$ and $\lambda _{SH}$ are real constants. For the SM sector, the
electroweak symmetry of the SM $G_{EW}=\mbox{SU(2)}_{L}\otimes \mbox{U(1)}%
_{Y}\subset G_{SM}$ is indeed broken by a non-vanishing vacuum expectation
value (VEV) of the Higgs field $\left\langle H\right\rangle =\left\langle
h^{0}\right\rangle $ $\sim 10^{2}GeV$ \cite{19}. For the $\mbox{U(1)}_{X}$
symmetry of the conserved number $X$, a close computation reveals that the
breaking is realized if $-(\mu _{H}^{2}+\mu _{S}^{2}+\left( 2\lambda
_{H}+\lambda _{SH}\right) \left\langle H\right\rangle ^{2})/2\lambda
_{S}+\lambda _{SH}>0$. In such a case, the scalar field $S$ develops a real
VEV given by
\begin{equation}
\left\langle S\right\rangle =\sqrt{\frac{-\left[ \mu _{H}^{2}+\mu
_{S}^{2}+\left( 2\lambda _{H}+\lambda _{SH}\right) \left\langle
H\right\rangle ^{2}\right] }{2\lambda _{S}+\lambda _{SH}}.}
\label{9}
\end{equation}%
The corresponding mass   reads as
\begin{equation}
m_{S}=\sqrt{-(\mu _{S}^{2}+\lambda _{SH}\left\langle H\right\rangle ^{2})}.
\label{10}
\end{equation}%
In this view, the interaction of the scalar $S$ with the SM sector results
indirectly through its coupling with the Higgs sector described by the term $%
\lambda _{SH}|S|^{2}|H|^{2}$. It turns out that this term has  to be
very small $\lambda _{SH}\ll 1$. Otherwise, this would impact
seriously the Higgs decay channels with the rate expected in the SM
\cite{1,2,3}. In the present model, this can be interpreted as a
small mixing between the two scalars $H$ and $S$ ensuring a
negligible communication between the SM sector and the new
$\mbox{U(1)}_{X}$ symmetry. In this way, no relevant effects on the
SM gauge sector will take place.

At this stage, we can think about the physical meaning of the presumed
broken symmetry $\mbox{ U(1)}_{X}$. Because there is no place for a new
broken global symmetry in the SM, it is natural to look of a symmetry
associated particles beyond the SM, but whose existence is theoretically
supported. A likely case is that of spin-%
$\frac12$
massive MMs $m$. Thus, we assume now that the conserved quantum number $X$
corresponding to the symmetry $\mbox{U(1)}_{X}$ is the magnetic number (\ref%
{1}), that is

\begin{equation}
X\equiv M=nM_{D},  \label{11}
\end{equation}%
being the number of MMs $m$ minus the number of the their antiparticles $%
\overline{m}$ like
\begin{equation}
M=N_{m}-N_{\overline{m}}.  \label{12}
\end{equation}%
At this level, one should add one comment on this magnetic symmetry $%
\mbox{U(1)}_{M}$. If we consider the charged magnetic particles $m$,
together with the complex scalar field $S$, the magnetic quantum number $M$
of the charged SM fields being neutral under the $\mbox{U(1)}_{M}$ should be
zero. According to (\ref{5}) and (\ref{6}), we list, thus, a set of magnetic
charges of SM and the added fields
\begin{eqnarray}
M\left( q_{i}\right) &=&N_{m}\left( q_{i}\right) -N_{\overline{m}}\left(
q_{i}\right) =0,  \label{13} \\
M\left( \ell _{i}\right) &=&N_{m}\left( \ell _{i}\right) -N_{\overline{m}%
}\left( \ell _{i}\right) =0,  \label{14} \\
\text{ }M\left( H\right) &=&0,  \label{15} \\
M\left( m\right) &=&N_{m}\left( m\right) -N_{\overline{m}}\left( m\right) =1,%
\text{ }  \label{16} \\
M\left( S\right) &\neq &0,  \label{17}
\end{eqnarray}
where the magnetic charge $M\left( S\right)$ of the scalar field $S$
could be specified in order  to elaborate consistently the present
physical model.  Giving the scalar field $S$ an arbitrary magnetic
quantum number $M(S)=nM_D$, the most general renormalizable terms
involving the added fields, that can be supplemented to the SM
lagrangian, are
\begin{equation}
\zeta _{MM}=\overline{m}\gamma ^{\mu }\partial _{\mu }m-\lambda
_{Sm}\frac{\overline{S}S\overline{m}m}{M_s}+h.c..  \label{18}
\end{equation}%
where $M_s$ denotes an underlying high physical  scale. $\lambda
_{Sm}$ is real coupling constant.   These relevant terms describe
the MMs propagation and the communication with the scalar field $S$
with a coupling constant $\lambda _{Sm}$. Using the non-zero VEV of
the scalar field $S$ given in (\ref{9}), the mass of MM particles is
\begin{equation}
m_{m}=-\frac{\lambda_{Sm}}{M_s}{\frac{\left[ \mu _{H}^{2}+\mu
_{S}^{2}+\left(
2\lambda _{H}+\lambda _{SH}\right) \left\langle H\right\rangle ^{2}\right] }{%
2\lambda _{S}+\lambda _{SH}}}.  \label{19}
\end{equation}%
In this route, it is noted MMs mass is indeed  determined by the breaking scale (\ref%
{9}) of the magnetic symmetry $\mbox{U(1)}_{M}$ and  the ${M_s}$
mass scale. It is recalled  that their masses could be largely heavy
$m_{m}\gg M_{SM}$. This could lead to the idea that there is no hope
of producing them in any foreseen accelerator. However, a best hope
is observing them in cosmic rays of the Universe.

\section{Searches for MMs}

It has been suggested that the detection of MMs might be either indirectly
via their annihilation product searches or directly by studying their
communication with ordinary particles within the detector. It should be
noted that one important difference between MMs and the most other SM
particles resides on the fact that the ligther MMs, if they exist, would be
absolutely stable. Whereas, they could disappear only if they meet other MMs
of opposite charges. In this way, they would self-annihilate and produce a
burst of lighter SM spectrum vie the following process
\begin{equation}
m\overline{m}\rightarrow \mbox{SM}.  \label{20}
\end{equation}%
By virtue of the magnetic charge $M$ conservation (\ref{11}), MMs would
always be produced in north--south pairs from SM particles. This can be
happened if the collision energy is higher than the combined mass of the two
MMs, that is $E_{Collision}\geqslant 2m_{m}$. However, if MMs are
sufficiently light, it is envisaged that they would be produced in
experiments using a process occurring especially in SM particle collisions.
Concretely, we could have the following scheme
\begin{equation}
q\overline{q}/\text{ }e^{-}e^{+}\rightarrow \gamma ^{\ast }\rightarrow m%
\overline{m}.  \label{21}
\end{equation}%
These kinds of searches are still going on at the powerful accelerators,
especially LHC with a maximum collision energy of the order of $E_{p%
\overline{p}}\sim 10^{4}GeV$. Whilst, the main emphasis of MMs
searches is clearly concentrated on high-mass low velocity cosmic
MMs. Assuming that the $\mbox{U(1)}_{M}$ symmetry breaking scale is
high $\left\langle S\right\rangle \gg \left\langle H\right\rangle
\sim 10^{2}GeV$, this would make the MMs heavy $m_{m}\sim
\left\langle S\right\rangle^2/M_s $ and far to be produced at any
accelerator if $M_s$ is not so huge,  that is $M_s\sim M_{planck}$.
In the case where MMs are as heavy as expected, the best hope is
observing them in cosmic rays, inspected by the actual telescopes,
for instance Icecube  and ANTARES \cite{3,4,5,6}. Using the fact
that in the present Universe no process occurring is sufficiently
energetic to produce MMs, any MMs today must be produced when higher
energies were available, in the very early Universe. As the universe
cooled down, the MMs density would have initially decreased by pair
annihilation given in (\ref{20}). Once they were sufficiently
reduced, the MMs would no longer be able to find annihilation
partners and they would be survived indefinitely. This  means that
the abundance of MMs is a cosmological issue. According to
astrophysical arguments, the flux of MMs in cosmic rays is probably
quite small. Moreover, the heavy MMs traversing the Earth are likely
to be moving relatively slowly, with a typical speed of the order of
$10^{-3}c$. Experimentally, the detection of such rare and slow
particles is considered as a challenging problem of modern physics.

In the intergalactic medium, acceleration of MMs by magnetic fields
can give them a kinetic energy up to $10^{14}GeV$ to reach the
terrestrial detectors \cite{3,4,5,6}. Considering this data and
using (\ref{19}), the kinetic energy of MMs could be written as
\begin{equation}
k_{m}\simeq -\frac{\lambda_{Sm}}{2M_s}{\frac{\left[ \mu _{H}^{2}+\mu
_{S}^{2}+\left(
2\lambda _{H}+\lambda _{SH}\right) \left\langle H\right\rangle ^{2}\right] }{%
2\lambda _{S}+\lambda _{SH}}}10^{-6}\lesssim 10^{14}GeV,  \label{22}
\end{equation}%
which, for $\lambda _{Sm}\sim \lambda _{S}\leq 1$, corresponds to the MMs
mass bound
\begin{equation}
m_{m}=-\frac{\lambda_{Sm}}{M_s}{\frac{\left[ \mu _{H}^{2}+\mu
_{S}^{2}+\left(
2\lambda _{H}+\lambda _{SH}\right) \left\langle H\right\rangle ^{2}\right] }{%
2\lambda _{S}+\lambda _{SH}}}\lesssim 10^{8}GeV.  \label{23}
\end{equation}%
This MMs mass bound allows one to approach the breaking scale of the
magnetic symmetry $\mbox{U(1)}_{M}$ as well as the mass of the
associated scalar. Concretely,  we have
\begin{equation}
m_{S}\sim \left\langle S\right\rangle =\sqrt{\frac{-\left[ \mu _{H}^{2}+\mu
_{S}^{2}+\left( 2\lambda _{H}+\lambda _{SH}\right) \left\langle
H\right\rangle ^{2}\right] }{2\lambda _{S}+\lambda _{SH}}}\gtrsim 10^{8}GeV.
\label{24}
\end{equation}%
It follows that the breaking of the magnetic symmetry
$\mbox{U(1)}_{M}$ and the mass of the scalar field $S$ overtake, as
expected, the SM scale. According to the MMs higher mass bound
(\ref{23}), however, the hope for a lighter MM mass to be seen in
current powerful accelerators remains plausible. In this picture,
the MMs can appear at the moment of corresponding to the spontaneous
breaking of the a high-scale magnetic symmetry of a large symmetry
group  $G_s$ into subgroups, one of which is the $U(1)_{Q}$ of
electromagnetism. In a group theory language, we end up with the
following scale transitions

\begin{equation}
G_s\overset{M_s}{\rightarrow }G_{SM}\otimes \mbox{U(1)}_{M}%
\overset{\left\langle S\right\rangle}{\rightarrow }\\G_{SM}%
\overset{\left\langle H\right\rangle}{\rightarrow }%
SU(3)_{C}\otimes U(1)_{Q}.  \label{25}
\end{equation}%
Unfortunately, the too many free parameters involved in the present study
and the approximations of some of them do not allow a precise result and
profound predictions, whether the MMs in this model do or do not match the
theoretical and observational constraints.

\section{Summary and outlook}
In this work, motivated by recent high energy detector activities,
we have attempted to give as an origin of MMs a new
$\mbox{U(1)}_{M}$ symmetry being hidden through a high-scale
spontaneous symmetry breaking. A new scalar field $S$ having a
non-zero VEV has been introduced to break such a new abelian
symmetry. The conserved quantum number, specified as the magnetic
number $M$, has been associated with the MMs, being charged together
with the scalar field $S$ under this magnetic symmetry unlike the SM
particles. Under this suggestion, the communication between the MMs
sector and the SM sector is ensured indirectly by such a scalar
field $S$ interacting with the
electroweak SM Higgs field $H$ through a weak mixing, governed by $%
\lambda_{SH}\ll 1$. Breaking the magnetic symmetry and assigning the
magnetic numbers to the spectrum of the extended-SM, we have given
allowed interaction terms of the new fields. Then, we have given an
explicit statement of the corresponding MMs mass in terms of the
involved scales in the model. After deriving the mass of MMs, we
have discussed their searches by means of direct or indirect
detections. In particular, if there is a hope for detecting of the
lightest MM, it would be absolutely stable. Thus, they must be seen
in the Universe as cosmic relics after the end of their
self-annihilating into lighter SM particles. In current large
accelerators, these lighter MMs would always be produced in
north-south pairs from SM particles as stated by magnetic charge $M$
conservation. Then, we have given an important attention to heavy
MMs, as incited by most GUT theories. Such MMs must have to be
produced at the very early Universe when higher energies were
available and would have survived today as cosmic relics, to be
detected in current high-energy particle telescopes. According to
the theoretical data and the astrophysical constraints on the
typical speed and likely kinetic energy of heavy MMs, we have
approached the involved scales associated with the magnetic symmetry
$\mbox{U(1)}_{M}$ by bounding its breaking scale $\sim \left\langle
S\right\rangle \gtrsim 10^{8}GeV$ and the mass of MMs $m_{m}\lesssim
10^{8}GeV$. It has observed that this mass is weighted   by the
underling mass scale $M_s$ supporting the  existence  of MMs. Though
these mass scales lie beyond the present accessible scale $\sim
TeV$, the expectancy of lighter MMs to be seen in the running
powerful accelerators remains reasonable. \newline
{\color{white}{..}}
 This work opens up for further studies. In connection with the compactification of string
theory, one could consider stringy models with several scalar fields
$S_i$. These scalars  could be associated with R-R and NS-NS fields
on 2-cycles of the Calabi-Yau manifolds. The number of such fields
which can be given in terms the size and the shape deformations of
the Calabi-Yau manifold can provide a statistical analysis. However,
the corresponding physics deserves a better understanding. We hope
to report elsewhere on these open questions. \newline
 {\color{white}{...}} In spite of their elusiveness, MMs still attract much of our attention and push us to
unveil some of their hidden properties and explore others. Both
theorists and experimenters will continue to follow MMs without tiredness.%
\newline
\newline
\textbf{Acknowledgement}: The authors would like to deeply thank
their family for all kind of supports and encouragement.

\end{document}